\documentstyle[twoside,fleqn,espcrc2,epsfig]{article}

\newcommand{\eqref}[1]{(\ref{#1})}

\newcommand{\AmS}{{\protect\the\textfont2
  A\kern-.1667em\lower.5ex\hbox{M}\kern-.125emS}}

\title{ \hfill  {\small HLRZ1999\_41}\\
  Scaling of magnetic monopoles
  in the pure compact QED
}
\author{J.~Jers{\'a}k, T.~Neuhaus, and H.~Pfeiffer\address{Institut f{\"u}r
Theoretische Physik E, RWTH Aachen, Germany}}
      
\begin{document}

\begin{abstract}
In the pure U(1) lattice gauge theory with the Villain action we find that
the monopole mass in the Coulomb phase and the monopole condensate in the
confinement phase scale according to simple power laws.  This holds outside
the coupling region in which on finite toroidal lattices the metastability
phenomena occur. A natural explanation of the observed accuracy of the
scaling behaviour would be the second order of the phase transition between
both phases in the general space of couplings not far away from the Villain
action.

\end{abstract}

\maketitle

\section{MOTIVATION}

The phase transition between the confinement and Coulomb phases of the
strongly coupled pure U(1) lattice gauge theory (pure compact QED) remains to
be puzzling. For the extended Wilson and Villain action, the presence of the
two-state signal on finite lattices has been recently confirmed. On the other
hand, a scaling behaviour of various bulk quantities and of the gauge-ball
spectrum consistent with a second order phase transition and universality has
been observed outside the narrow region in which the two-state signal occurs.
(See ref.~\cite{JeNe99} for references.)

Because the order of the phase transition for these actions is unknown, the
extrapolation of these phenomena to the thermodynamic limit is uncertain.
However, even if the scaling behaviour is only a transient phenomenon, it
indicates that there is a region of the phase diagram described by an
interacting effective field theory. It is of interest to investigate the
properties of such a theory even if it is ``only'' effective, as effective
theories are useful in physics. Here we address the question whether such a
theory includes monopole degrees of freedom.

\section{MAIN RESULTS}

We have observed scaling of some observables related to the magnetic monopoles
in the pure compact QED with Villain action.

In the Coulomb phase we find at various values of the coupling $\beta$ a very
clean exponential decay of the monopole correlation function in a large range
of distances. This demonstrates the dominance of a single particle state in
this correlation function, the monopole, whose mass we determine. Due to its
Coulomb magnetic field, the monopole mass strongly depends on the finite
lattice size. However, we find \cite{JeNe99} that it can be reliably
extrapolated to the infinite volume.

\begin{figure}[t]
\begin{center}
  \psfig{file=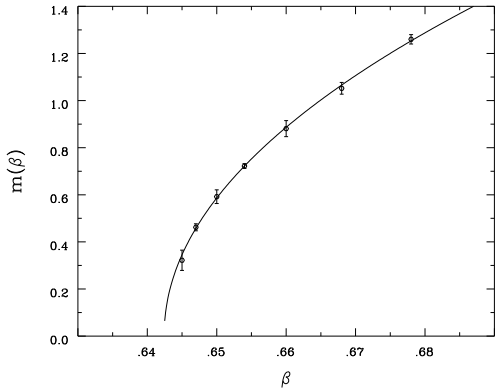,angle=90,width=\hsize}
  \caption{
    \label{monoscale} Scaling behaviour of the monopole mass extrapolated 
    to $L = \infty$ with $\beta$. The curve corresponds to the power 
    law~(\ref{m_infty}).}
\end{center}
\end{figure}

The scaling behaviour of the extrapolated monopole mass $m_\infty$ at
the phase transition follows a simple power law (fig.~\ref{monoscale})
\begin{equation}
       m_\infty(\beta)=a_m{(\beta-\beta_c^{\rm Coul})}^{\nu_{\rm m}},
       \label{m_infty}
\end{equation}
with the critical exponent
\begin{equation}
       \nu_{\rm m}=0.49(4).
       \label{nu_m}
\end{equation}
The inverse mass achieves at least the magnitude of three lattice spacings.

\begin{figure}[t]
\begin{center}
  \psfig{file=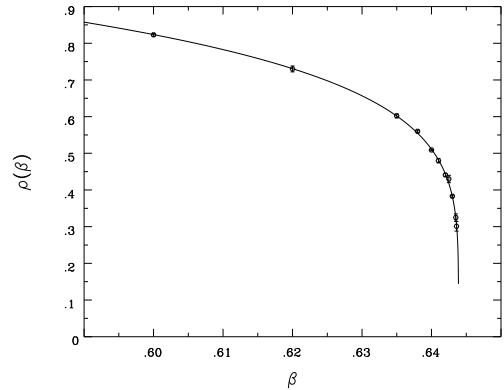,angle=90,width=\hsize}
  \caption{
    \label{condfig} Scaling behaviour of the monopole condensate extrapolated 
    to $L = \infty$ with $\beta$. The curve corresponds to the power
    law~(\ref{condscale}).}
\end{center}
\end{figure}

The monopole condensate in the confinement phase shows a much weaker
$L$ dependence. Its value extrapolated to the infinite
volume, $\rho_\infty$, scales with the power law
\begin{equation}
\label{condscale}
  \rho_\infty=a_\rho{(\beta_c^{\rm conf}-\beta)}^{\beta_{\rm exp}},
\end{equation}
whith the magnetic exponent
\begin{equation}
\label{magneticexp}
  \beta_{\rm exp}=0.197(3).
\end{equation}
As shown in fig.~\ref{condfig} the function (\ref{condscale}) describes
extremely well the data in a broad interval and the scaling behaviour of the
condensate is thus well established.

The superscripts ``Coul'' and ``conf'' indicate that the corresponding
values of $\beta_c$ have been determined by the power law fits using
data only from one phase. Their values are
\begin{equation}
  \beta_c^{\rm Coul}=0.6424(9)
\end{equation}
and
\begin{equation}
  \beta_c^{\rm conf}=0.6438(1).
\end{equation}
Both values are consistent within two error bars. 

Further results and technical details of our calculations are published in
\cite{JeNe99}. We have adopted the methods of ref.~\cite{PoWi91}. A quantity
related to $\rho_\infty$ has been studied also in refs.~\cite{DiLu98}.

\section{INTERPRETATION OF RESULTS}

The monopole mass in the Coulomb phase scales with the same Gaussian exponent
$\nu_m$ which is also observed for the scalar gauge ball. This holds at least
until the inverse mass of the latter achieves five lattice spacings. This
implies that if one chooses the scalar gauge ball to become massless while the
other gauge balls, whose $\nu$ is about 1/3, have finite non-vanishing masses,
the monopoles will be massless and therefore important. Even if the scalar
mass is chosen finite nonzero, and other gauge balls thus decouple, the
monopoles stay present.  Therefore the effective field theory would include
monopole degrees of freedom, being thus a very interesting abelian gauge
theory.  This may be a sufficient motivation for further investigation of
compact QED by the lattice community.

\begin{figure}[t]
\begin{center}
  \psfig{file=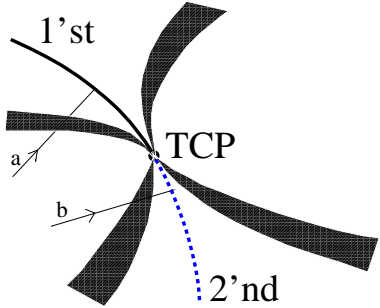,angle=270,width=5cm}
  \caption{
    \label{tcp}Generic cross-over regions in the vicinity of a tricritical
    point TCP.}
\end{center}
\end{figure}

Now let us try to interpret our results from the point of view of Statistical
Mechanics.  The coexistence of first and second order phenomena is a typical
property of tricritical points (TCP) \cite{JeNe85}. As indicated schematically
in fig.~\ref{tcp}, in their vicinity crossover regions (shaded) separate
regions of different behaviour even in the thermodynamic limit.

Approachig the phase transition along the path {\bf a} may reveal first a
second-order-like behaviour determined by the tricritical point, and only very
close to the phase transition the presence of the two-state signal shows
up. In finite systems, a two-state signal can appear even at the end of the
path {\bf b}.

The observed properties of the compact QED with various actions can be
explained by assuming the existence of a tricritical part of the mani\-fold
separating the confinement and Coulomb phases in the multidimensional space of
possible couplings. Thus under this hypothesis a genuine continuum limit of
the compact QED would exist.

Such a manifold may, but does not need to include the couplings which have
been already used for the investigation of compact QED.  Therefore, the search
for this manifold may require an introduction and investigation of new types
of coupling terms. As the monopoles are relevant, the space of generalized
couplings in which the TCP is to be located, is likely to include the monopole
degrees of freedom. Their influence on the transition has been studied in
refs. \cite{BaShr}. A possible TCP in this context has been discussed by
Kleinert \cite{Kl83}

However, even if the finding of such a manifold may be challenging, the
indications for its existence are remarkable: (i) the clean scaling behaviour
like that of the monopole observables, and (ii) several universal phenomena in
some intervals of couplings close to the phase transition points.  In fact,
these properties allow to investigate the corresponding continuum limit
without an actual localization of the tricritical points.

\begin{figure}[t]
\begin{center}
  \psfig{file=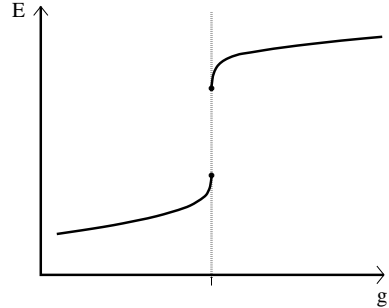,angle=270,width=5cm}
  \caption{
    \label{hybrid}Generic hybrid phase transition with simultaneous first 
    and second order behaviour.}
\end{center}
\end{figure}

Another scenario is that the coexistence of first and second order
phenomena is due to a rare, but not impossible hybrid situation depicted in
fig.~\ref{hybrid}. A continuum limit would then exist in spite of the
latent heat present in the thermodynamic limit.

\section*{ACKNOWLEDGEMENTS}

We thank J. Cox and U.-J. Wiese for discussions and suggestions. Computations
have been performed at NIC J\"ulich (former HLRZ J\"ulich).


\end{document}